\documentclass{aastex}
\usepackage{emulateapj5}
\usepackage{onecolfloat5}
\usepackage{apjfonts}
\usepackage{epsfig} 

\def\myputfigure#1#2#3#4#5%
{\vskip#5pt\makebox[0pt]{\hskip#2in
\includegraphics[width=#3\textwidth]{#1}}\vskip#4pt\hfill}

\def\gsim{\;\rlap{\lower 2.5pt
 \hbox{$\sim$}}\raise 1.5pt\hbox{$>$}\;}
\def\lsim{\;\rlap{\lower 2.5pt
   \hbox{$\sim$}}\raise 1.5pt\hbox{$<$}\;}
\newcommand{\be}{\begin{equation}}
\newcommand{\beq}{\begin{equation}}
\newcommand{\ba}{\begin{eqnarray}}
\newcommand{\ee}{\end{equation}}
\newcommand{\eeq}{\end{equation}}
\newcommand{\ea}{\end{eqnarray}}

\newcommand{\hs}{\hspace{1mm}}

\newcommand{\lya}{Ly$\alpha \hspace{1mm}$}
\newcommand{\ha}{H$\alpha \hspace{1mm}$}

\begin{document}
\twocolumn[

\title{Ly$\alpha$ Radiation From Collapsing Protogalaxies II:\\Observational Evidence for Gas Infall.}

\author{Mark Dijkstra\altaffilmark{1,3}, Zolt\'an Haiman\altaffilmark{1} \& Marco Spaans\altaffilmark{2}}

\affil{$^1$Department of Astronomy, Columbia University, 550 West 120th Street, New York, NY 10027}

\affil{$^2$Kapteyn Astronomical Institute, P.O. Box 800, 9700 AV Groningen, The Netherlands}

\affil{$^3$School of Physics, University of Melbourne, Parkville, Victoria 3010, Australia}

\vspace{0.23cm}
\vspace{-0.5\baselineskip}
 
\begin{abstract}
We model the spectra and surface brightness distributions for the
Ly$\alpha$ radiation expected from protogalaxies that are caught in
the early stages of their assembly.  We use the results of a companion
paper to characterize the radiation emerging from spherically
collapsing gas clouds. We then modify these spectra to
incorporate the effect of subsequent resonant scattering in the
intergalactic medium (IGM).  Using these models, we interpret a number
of recent observations of extended Ly$\alpha$ blobs (LABs) at high
redshift.  We suggest, based on the angular size, energetics, as well
as the relatively shallow surface brightness profiles, and
double-peaked spectra, that several of these LABs may be associated
with collapsing protogalaxies.  We suggest two follow-up observations
to diagnose the presence of gas infall.  High S/N spectra of LABs
should reveal a preferential flattening of the surface brightness
profile at the red side of the line. Complementary imaging of the
blobs at redshifted H$\alpha$ wavelengths should reveal the intrinsic
Ly$\alpha$ emissivity and allow its separation from radiative transfer
effects. We show that Ly$\alpha$ scattering by infalling gas can
reproduce the observed spectrum of Steidel et al's LAB2 as accurately
as a recently proposed outflow model. Finally, we find similar
evidence for infall in the spectra of point--like Ly$\alpha$
emitters. The presence of scattering by the infalling gas implies that
the intrinsic Ly$\alpha$ luminosities, and derived quantities, such as
the star--formation rate, in these objects may have been
underestimated by about an order of magnitude.
\end{abstract}

\keywords{galaxies: formation -- galaxies: halos -- quasars: general -- radiative transfer -- cosmology: theory -- intergalactic medium}]


\section{Introduction}
\label{sec:intro}

Copious Ly$\alpha$ emission may be associated with the early stages of
the formation of galaxies. The epoch before the onset of significant
star formation in the history of an individual galaxy, hereafter
referred to as the `protogalactic' stage, may be accompanied by the
release of the gravitational binding energy through Ly$\alpha$ cooling
radiation \citep{Haiman00,Fardal01}. The extended Ly$\alpha$ ``fuzz''
surrounding these protogalaxies may be strongly enhanced in the
presence of a central ionizing source, such as a quasar
\citep{Haiman01,Haiman04w}, as evidenced by a handful of recent
observations \citep{Bunker03,Weidinger04,Weidinger05,Christensen06}.  As the
number of known high redshift Ly$\alpha$ emitters, both point--like
\citep[e.g.][]{rhoads00,Rhoads01,Ajiki03,Kodaira03,Hu04,Rhoads04,Ouchi05}
and extended \citep[e.g][]{Steidel00,matsuda04,Wilman05,Dey05,Nilsson05,Matsuda06} grows, there is an increasing need for diagnosing the presence of any large--scale
gas infall.

In \citet[][hereafter paper I]{PaperI} we presented Monte Carlo
simulations of the transfer of Ly$\alpha$ photons through spherically
symmetric models of neutral collapsing gas clouds. These models are
intended as simplified descriptions of protogalaxies in the process of
their assembly.  We computed the spectra and surface brightness
profiles emerging from clouds with a range of properties.  For clouds
without any associated ionizing source, generating Ly$\alpha$ photons
through cooling radiation alone, we explored two extreme models. In
the `extended models', the gas continuously emits a significant
fraction of its gravitational binding energy in the Ly$\alpha$ line as
it navigates down the dark matter potential well. In the ``central''
models, \lya photons are generated only at the center of the
collapsing cloud. We then studied how the results of these
calculations were affected by the presence of a central ionizing
source. The main results that are relevant to the present paper can be
summarized as follows:

(i) In the absence of an ionizing source (i.e. stars or a quasar) the emergent Ly$\alpha$ spectrum is typically double--peaked and asymmetric. Transfer of energy from the collapsing gas to the Ly$\alpha$ photons -together with a reduced escape probability for photons in the red wing- causes the blue peak to be significantly enhanced, which results in an effective blueshift of the Ly$\alpha$ line. This blueshift may easily be as large as $\sim 2000$ km s$^{-1}$. The total blueshift of the line increases with the total column density of neutral hydrogen of the gas and with its infall speed. The prominence of the red peak is enhanced towards higher redshift, lower mass and lower infall speeds. If detected, the shift of the red peak relative to the line center potentially measures the gas infall speed.

(ii) In the ``central'' models, the Ly$\alpha$ radiation emerges with
a steeper surface brightness profile, and with a larger blueshift,
than in the ``extended'' models.

(iii) A steepening of the surface brightness distribution towards bluer
wavelengths within the Ly$\alpha$ line is indicative of gas infall.

(iv) In clouds that are fully ionized by an embedded ionizing source,
the effective Ly$\alpha$ optical depth is reduced typically to $\lsim
10^3$. In these cases, the Ly$\alpha$ emission emerges with a nearly
symmetric profile, and its FWHM is determined solely by the bulk
velocity field of the gas. Its overall blueshift is reduced
significantly (and may vanish completely for very bright ionizing
sources). As a result, infall and outflow models can produce nearly
identical spectra and surface brightness distributions, and are
difficult to distinguish from one another.
 
For a detailed description of several other features of the emergent
Ly$\alpha$ radiation the reader is referred to paper I. The models
discussed in paper I are oversimplified, for a discussion on the 
uncertainties this may introduce the reader is referred to paper I, 
specifically section 8.2. The results presented in paper I generally ignored
the transfer through the intergalactic medium (IGM). Since the
IGM is optically thick at the Ly$\alpha$ frequency at $z\gsim 3$, to
allow comparison of these results with the observations, the
scattering of Ly$\alpha$ photons through the IGM needs to be
modeled. Note that we define the IGM to commence beyond the virial
radius of an object. The intrinsic spectrum refers to the
spectrum as it emerges at the virial radius. In the present paper, we
adopt a simple model of the IGM gas density and velocity distribution
expected near protogalaxies, based on the models of \citet{Barkana04}.
Combining the results obtained in paper I with the scattering in the
IGM, we compare our models to several observations in the literature.
We focus on interpreting those extended Ly$\alpha$ blobs (LABs) that
have no continuum source, or surround a radio quiet quasar.  While
LABs are known to exist around radio loud quasars, they are
attributable to known jets and supernova--driven outflows, and we do
not attempt to model them here \citep[e.g.][]{Hu91,Wilman00}.  We also
compare our results with the recently discovered high redshift ($z >
5$) Ly$\alpha$ emitters, which generally appear point--like.  Despite
the over--simplified nature of our model of the protogalactic gas and
the IGM, we successfully reproduce several observed Ly$\alpha$
spectral features.

The rest of this paper is organized as follows.
In \S~\ref{sec:IGM}, we describe our model of the IGM and discuss its impact
on our results from paper I.
In \S~\ref{sec:observations}, we compare our model predictions with
the properties of several known high--redshift Ly$\alpha$ emitting
sources.
In \S~\ref{sec:discussion}, we discuss our results and their
 cosmological relevance. We also discuss the possibility of imaging the protogalaxies in the H$\alpha$. Since this line should not suffer from attenuation
in the IGM, it may have a surface brightness comparable to (or even
exceeding) that of the Ly$\alpha$ line.
Finally, in \S~\ref{sec:conclusions}, we present our conclusions and
summarize the implications of this work.
The parameters for the background cosmology used throughout this paper
are $\Omega_m=0.3$, $\Omega_{\Lambda}=0.7$, $\Omega_b=0.044$, $h=0.7$,
based on \citet{Spergel03}.

\section{The Impact of the IGM}
\label{sec:IGM}

\subsection{Impact of a Static IGM}
 In the simplest model, employed occasionally in paper I, 
Ly$\alpha$ photons are injected into a statistically 
uniform IGM, following Hubble expansion. Photons with an original wavelength blueward of the Ly$\alpha$ line center redshift into resonance and may get scattered out of our line of sight. Ly$\alpha$ photons that are initially
redward of the line center are not scattered resonantly (scattering in
the red damping wing is not important in the present context, at
redshifts at which the IGM is highly ionized). We follow paper I and
define $x$ as a dimensionless frequency variable, which measures the offset
from the line center in units of the Doppler width, $x\equiv(\nu-\nu_0)/\Delta \nu_D$. Note that positive/negative values of $x$
correspond to blueshifted/redshifted Ly$\alpha$ photons.  The IGM 
then allows a transmission of a fraction $\langle \exp{(-\tau)}\rangle$
and $1$ with $x>0$ and $x \leq 0$ of photons emerging from the collapsing cloud, respectively. Here, $\langle \exp{(-\tau)} \rangle$ denotes
the mean IGM transmission as it has been determined observationally 
in studies of the Ly$\alpha$ forest. Table~\ref{table:IGM} shows $\langle \exp{(-\tau)}\rangle$ as a function of redshift. The second row shows $\sigma_f$, which is the reported measurement error ($1-\sigma$) in $f$. This Table shows that attenuation by the IGM is significant at $z\gsim 3$.
\begin{table}[t]
\small
\caption{Transmission of IGM}
\label{table:IGM}
\begin{center}
\begin{tabular}{lccccccc}
\tablewidth{1in}
z & 2.4 \tablenotemark{1} & 3.0 \tablenotemark{1}&3.8 \tablenotemark{1}& 4.5 \tablenotemark{2}& 5.2 \tablenotemark{2}& 5.7 \tablenotemark{2}& 6.0\tablenotemark{2} \\
\hline
$f \equiv \langle \exp{(-\tau)}\rangle$ & 0.8 & 0.68 & 0.49 & 0.25 & 0.09 & 0.07 & 0.004 \\
$\sigma_f$ &   &  &  &  & 0.02 & 0.003 & 0.003 \\
\hline
\multicolumn{8}{l}{(1) From \citet{McDonald01}}\\
\multicolumn{8}{l}{(2) From \citet{Songaila02,Fan02}}\\
\end{tabular}\\[12pt]
\end{center}
\end{table}

\subsection{Impact of a Dynamic IGM}
Protogalaxies are likely to reside in relatively massive halos,
corresponding to high--$\sigma$ density peaks in the IGM.  The
gravitational field of the halos causes material on large scales to
decouple from the Hubble flow and fall in toward the halo.
 \citet{Barkana04} has modelled the density
and velocity distribution of the IGM around nonlinear halos. In these
models, around the massive halos typically considered in paper I, 
infall may occur up to about $\sim 5$ virial
radii. Photons that escape the protogalaxy with a small redshift, $x
\lsim 0$, may then be resonantly scattered by the infalling gas.  As a
result, significant attenuation can extend into the red part of the
observed spectrum.  To give a quantitative example (see Figure 2 and 4
in \citealt{Barkana04}), for a dark matter halo with a total mass of
$10^{12}~{\rm M_{\odot}}$, the overdensity of the IGM is expected to
be $\delta \sim 20-30$ at the virial radius, and still $\delta \sim
2-3$ at $\sim 10$ virial radii from the object. The infall velocity is
equal to the circular velocity at the virial radius. The IGM is at
rest with respect to the virialized object approximately 5 virial
radii away. The radius at which the peculiar velocity of the IGM
vanishes (and is therefore comoving with the Hubble flow ) is 
further out, and is mass dependent.

\begin{figure}[t]
\vbox{ \centerline{\epsfig{file=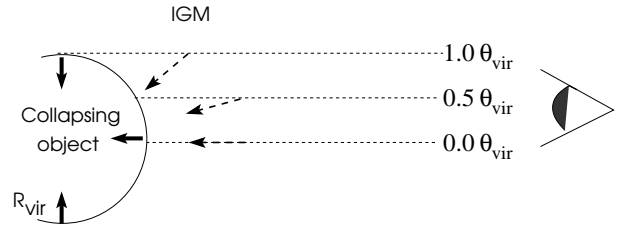,width=8.0truecm}}}
\caption[Drawing to illustrate the effect of a dynamical IGM.]{This
drawing illustrates the effect of the impact parameter on the IGM
transmission. The IGM is assumed to undergo spherically symmetric
infall. In this case, photons emerging from the edge of our object (in
the direction along the line of sight) see a projected version of the
IGM velocity field. }
\label{fig:igmscheme} 
\end{figure}

The impact of the IGM will be further modified by the proximity effect
near bright QSOs. For bright QSOs, the size of the local HII region
can extend beyond $\sim 5 R_{\rm vir}$, significantly reducing
resonant absorption from the dynamically biased part of the IGM
\citep{CenHaiman00, MadauRees00}.  In our models below, we allow for
the presence of such a cosmological HII region.  Finally, the impact
of the IGM can also vary from location to location, due to stochastic
density and ionizing background fluctuations. We will restrict our
study to study the mean effect of the IGM, and not investigate such
variations in the present paper.

\subsection{Modelling the Dynamic IGM}

To illustrate the impact of the IGM, we model the transmission around
a halo with a total mass $M_{\rm tot}=5.2 \times 10^{11}~{\rm
M_{\odot}}$ (corresponding to the mass of fiducial model {\it 1.} 
in paper I), in the presence or absence of a bright quasar. Note that for
increasing masses, the IGM would have increasing peculiar velocities,
and would affect a wider range of frequencies, extending further to
the red side of the line. We assume that the IGM is falling in with a
velocity profile $v_{\rm IGM}(r)$, and that the total number density
of hydrogen atoms (neutral+ionized) is $n_{\rm IGM}(r)$.  The optical
depth for a Ly$\alpha$ photon that enters the IGM at frequency $x$ is
given by
\begin{equation}
\tau_x=\int_{r_{\rm vir}}^{\infty} ds \hs n_{\rm IGM}(s) \hs x_{\rm
HI}(s)\hs\sigma_0 \hs\phi(x[s]),
\end{equation} 
where $x_{\rm HI}$ is the neutral fraction of hydrogen by number.  We
take density and velocity profiles which follow the curves shown in
Figures 2 and 4 in \citet{Barkana04}. We assume a gas temperature of
$T=10^4$ $^\circ$ K (as in paper I, appropriate for photoionized
gas). For an extended, spherically symmetric source, Ly$\alpha$
photons propagating along the line of sight ``see'' a velocity
component of the IGM that depends on their impact parameter. This is
illustrated in Figure~\ref{fig:igmscheme}. The angle $\theta_{\rm
vir}$ is the angle on the sky extended by the virial radius $r_{\rm
vir}$.  We adopt a standard expression for the virial radius, as a
function of halo mass and redshift, from \citet{BarkanaLoeb01}.
Photons that were received from the center of the object, $\theta=0$,
see the full magnitude of the IGM peculiar velocity vector. However,
photons that were received from $|\theta|>0$ see only the projection
of the infall velocity along the line of sight. As a result, the IGM
transmission depends on $\theta$,

\begin{figure}[t]
\vbox{ \centerline{\epsfig{file=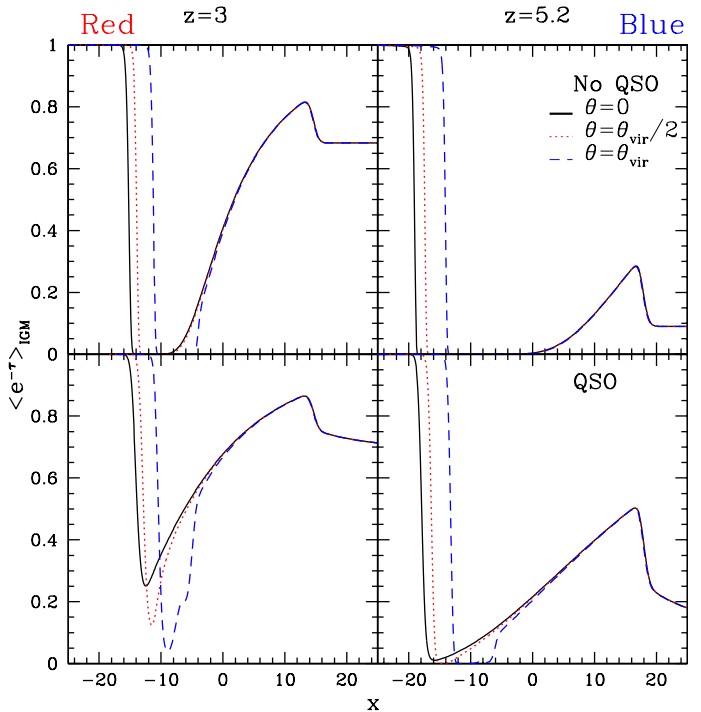,width=10.0truecm}}}
\caption[The \lya transmission through a dynamic IGM as a function of
frequency]{The transmitivity of a dynamic, infalling IGM 
surrounding a dark matter halo with total mass $M_{\rm tot}=5.2 \times 10^{11} M_{\odot}$ at $z=3$ ({\it Left Panels}) and $z=5.2$ ({\it Right Panels}). The  horizontal axis shows the frequency offset from the line center, in units of the Doppler width. The {\it upper/lower panels} show the transmission in
the absence/presence of a central ionizing source in the halo with
$L_{\rm ion}=5\times 10^{44}$ ergs s$^{-1}$, respectively. The {\it
solid--curve} is the transmission from the center of our object, and
is similar to the results found by \citet{Barkana042}. The {\it
dotted} and {\it dashed curves} show the transmission at an angular
impact parameter that corresponds to $0.5\theta_{\rm vir}$ and
$1.0\theta_{\rm vir}$, respectively ($\theta_{\rm vir}$ is the angular
size of the virial radius of the dark matter halo). The figure shows
the IGM transmission reaches a minimum redward of the true line center for all models. Furthermore, the IGM transmission is a function of impact parameter especially on the red side of the Ly$\alpha$ line ($-10\lsim x \lsim 0$), which may be important for a diffuse background source. }
\label{fig:igmtransmission} 
\end{figure}

\begin{equation}
\tau_x(\theta)=\sigma_0\int_{r_{\rm vir}}^{\infty}dr \frac{r\hs n_{\rm
IGM}(r)\hs x_{\rm H}(r)}{\sqrt{r^2-y^2}} \phi\Big{(}x-\frac{v_{\rm
IGM}(r)}{v_{\rm th}}\big{[}1-\frac{y^2}{r^2}\big{]}^{1/2}\Big{)},
\label{eq:tauigm}
\end{equation} 
where $y\equiv\theta\times[r_{\rm vir}/\theta_{\rm vir}]$, and the
integral over the line of sight $s$ has been converted to an integral
over the radial coordinate $r$ measuring the distance from the center
of the halo to points along the line of sight.

The final step in the calculation is to obtain the neutral fraction
$x_{\rm HI}(r)$.  As mentioned above, the mean transmission $\langle
\exp{(-\tau)} \rangle$ is known observationally.  Here the brackets
denote averaging over the probability distribution of overdensities,
arising from small-scale density fluctuations in the IGM. The
averaging is over a spatial scale corresponding to the spectral
resolution.  We make the simple Ansatz that $\langle \exp{(-\tau)}
\rangle$ equals $\exp{(-\tau_0)}$, where $\tau_0$ is the transmittivity
at the mean density and $x=0$,
\begin{equation}
\tau_0=\frac{\langle n_{\rm IGM}\rangle x_{\rm HI,0} v_{\rm
th}\sigma_0}{H(z)},
\end{equation} 
where in the last step we used $ds=v_{\rm th}dx/H(z)$ (see Appendix A
of paper I).  From this Ansatz, we
compute $x_{\rm HI,0}$, the neutral fraction at the mean density, and,
assuming photoionization equilibrium, we compute the background flux
$J_{\rm bg}$.  For gas at overdensity $\delta\neq 0$, the neutral
fraction then simply scales as $x_{\rm HI}(\delta)\propto (1+\delta)
J_{\rm bg}^{-1}$. In the absence of a quasar, $J_{\rm bg}$ is
constant, and in the presence of a quasar, we add the quasar's flux,
which scales as $J_{\rm Q}\propto r^{-2}$.  For a realistic density
probability distribution function (PDF), the transmission at the mean
density can differ appreciably from $\langle \exp{(-\tau)}
\rangle$. However, our simple treatment, by construction, will give
the correct mean transmittivity, and effectively assumes only that the
gas clumping factor $C=\langle \rho^2\rangle / \langle \rho\rangle^2$
due to small--scale inhomogeneities is independent of the overdensity
$\delta$ on larger scales (corresponding to a spectral resolution
element).  More elaborate treatments would incorporate a model for the
small--scale gas clumping. For example, \citet{Barkana042} model the
density PDF based on simulations. We find that our crude treatment
yields comparable results for the transmission curves they obtained in
their Figure 3 (especially when considering that their transmission
will be reduced further near the line center in the absence of a
quasar).

The IGM transmission we obtain is presented in
Figure~\ref{fig:igmtransmission} in four different cases. The
transmittivity is shown as a function of frequency $x$ for three
different impact parameters, $\theta=0$ ({\it solid--black curves}),
$\theta=0.5 \theta_{\rm vir}$ ({\it red--dotted curves}) and
$\theta=1.0 \theta_{\rm vir}$ ({\it blue--dashed curves}). The {\it
upper panels} show the transmission in the absence of any ionizing
sources (apart from the uniform background), while the models shown in
the {\it lower panels} include a source with a total ionizing
luminosity of $5\times 10^{44}$ ergs s$^{-1}$. We assumed a
quasar--like power--law spectrum for the source of the form $\propto
\nu^{-1.7}$. \cite{Barkana042} performed similar calculations for the
IGM transmission. The three main differences in the work presented in
this paper are (i) our simpler treatment of the small--scale gas
clumping, (ii) the possibility that no bright ionizing quasar is
present, and (iii) our spherically symmetric objects are spatially
extended, and the IGM transmission depends on the impact parameter.
The following can be inferred from Figure~\ref{fig:igmtransmission}:
\begin{itemize}
\item The IGM is most opaque at frequencies around the line center. Depending on the exact model, the IGM's opacity is at its maximum slightly redward of the true line center. The reduced IGM transmission extends to
frequencies corresponding to $\sim v_{\rm circ}$, as already concluded
by \citet{Barkana03,Barkana042};
\item In the absence of an ionizing quasar, the transmission is reduced
to $\sim 0$ for a significant range of frequencies \citep[whereas the
transmission never drops to 0 in][ due to the presence of a bright
quasar in their models]{Barkana042}. This range depends on the
redshift, with attenuation decreasing at lower $z$, as shown by the
difference between the left and right panels.  In the presence of the
ionizing source, the attenuation is significantly reduced, as shown by
comparing the upper and lower panels.
\item The IGM transmission is a function of impact parameter. The frequency range with strong attenuation diminishes at larger impact parameters, due to the smaller projected peculiar velocity of the IGM.

\end{itemize}

\subsection{Impact of a Dynamic IGM on the Results from Paper I.}

\begin{figure}[ht]
\vbox{ \centerline{\epsfig{file=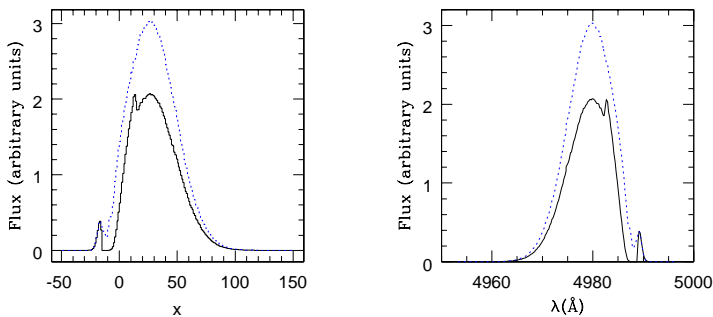,width=10.0truecm}}}
\caption[]{The impact of the IGM on the spectrum from our fiducial
{\it model 1}. The {\it left panel} shows the intrinsic
({\it blue--dotted line}) and observable ({\it black--solid line})
spectra. The IGM transmittivity from the {\it upper left panel} of
Fig.~\ref{fig:igmtransmission} is taken. In the {\it right panel}, the
spectrum is shown as a function of observable wavelength. The figure shows
that the IGM suppresses the blue peak and enhances the prominence of the dip between the red and blue peak.}
\label{fig:fid.igm} 
\end{figure}

First, we show how the IGM modifies the spectrum emerging from the
fiducial {\it model 1} from paper I. The {\it left panel} of
Figure~\ref{fig:fid.igm} shows the flux (in arbitrary units) as a
function of frequency, $x$, for the intrinsic ({\it
blue--dotted line}) and observable ({\it black--solid line})
cases. Since {\it model 1.} concerns the collapse of a neutral
hydrogen cloud (i.e. no embedded ionizing source is present), the IGM
transmittivity from Figure~\ref{fig:igmtransmission} for the `no qso'
case ({\it upper panels}) is taken. For this plot, the redshift of the
Ly$\alpha$ source was assumed to be $z=3$. In the {\it right panel},
the flux is shown as a function of observable wavelength.
The main effects of the IGM here are to suppress the blue peak and to
enhance the prominence of the dip between the blue and red peak. We
note that the dip remains at the same location. For the spectra
emerging from the central models, the photons emerge with a larger
average blue shift (with practically no photons redward of the line
center), and for these models the main effect of the IGM is to
suppress the total Ly$\alpha$ flux.

Second, we discuss how the IGM modifies the surface brightness profiles.
In the absence of a quasar (or any other ionizing source), the IGM becomes more transparent with $\theta$ for photons in the red peak. The result of this is that the surface brightness profiles of the reddest photons ($x \lsim -10$) appears flatter than those calculated in paper I. This would strengthen the result that for neutral collapsing protogalaxies (i.e. in the absence of an ionizing source), the reddest \lya photons appear less centrally concentrated than the
bluest \lya photons (see Figure 10 in paper I).

In the presence of a quasar (or any other ionizing source), the IGM also
flattens the surface brightness profile of the reddest photons ($x \lsim -10$).
 However, the transmission of the IGM can 
{\it decrease} with increasing $\theta$ for a range of
frequencies, $ 0 \lsim x \lsim -10$, e.g. at $x\approx -10$ in the
lower panels of Fig.~\ref{fig:igmtransmission}.  This is because at
larger impact parameters, there can be a large effective column of
absorbing material at projected velocity smaller than the maximum IGM
infall speed. For photons in this frequency range, the IGM steepens the 
surface brightness profiles.

The exact dependence of the IGM transmission on frequency and impact parameter in the presence (and absence) of an ionizing source depends strongly on the exact assumed density and velocity field of the gas in the IGM. Furthermore, the IGM is unlikely to collapse in a spherically symmetric fashion. This introduces
uncertainties to the angular dependence of the IGM transmission whose
modeling is beyond the scope of this paper. Despite these uncertainies, the main impact of the IGM on the results obtained in paper I is clear:

\begin{itemize}

\item The result that a steepening of the surface brightness profile towards bluer wavelengths within the Ly$\alpha$ line is a useful diagnostic for gas infall, is not affected by the IGM. 

\item The result that infall and outflow models can produce nearly identical spectra, may be affected by the IGM. The reason is that the IGM transmission is asymmetric around $x=0$. Below (\S~\ref{sec:LAB2}), we show that scattering in the IGM may cause emission lines that have identical profiles, but differ only by a small ($\sim$ several Doppler widths) frequency shift (caused by either inflow or outflow), may have somewhat different observed values of $[B]/[R]$ (defined in paper I as the ratio of the number of photons in the blue and red peak).

\item All other results relating to (i) the net blueshift of the blue peak, (ii) FWHM of the blue peak, (iii)  the location of the red peak (if any), (iv) the ratio $[B]/[R]$ and its dependence on mass, gas infall velocity and redshift are not affected significantly by the IGM.

\end{itemize}

With this knowledge at hand, and keeping in mind that a dynamic IGM may produce
an absorption feature slightly redward of the true Ly$\alpha$ line center, we interprete several observations of known Ly$\alpha$ emitters below.

\section{Comparison with Observed Lyman $\alpha$ Sources}
\label{sec:observations}

It will be useful to explicitly recall a result from
paper I, in which we found the total observable
Ly$\alpha$ cooling radiation flux from an object of mass $M_{\rm
tot}$, collapsing at redshift $z_{\rm vir}$, to be approximately
\begin{eqnarray}
\frac{f_{{\rm Ly}\alpha}}{10^{-18}}\approx 2 \hs f_{\alpha} \hs
\Big{(} \frac{M_{\rm tot}}{10^{11}M_{\odot}}\Big{)}^{\frac{5}{3}}
\Big{(} \frac{v_{\rm amp}}{v_{\rm circ}}\Big{)}\Big{(}
\frac{2-\alpha}{1.75}\Big{)}^{1.2}\times \nonumber \\ \left\{
\begin{array}{ll} \ \times (\frac{5}{1+z_{\rm vir}})^{1.75}& z \leq
4.0\\ \\ \ \times (\frac{5}{1+z_{\rm vir}})^{8.3} & z > 4.0
\end{array} \right.  \hs \frac{{\rm ergs}}{{\rm sec}\hs{\rm cm}^2},
\label{eq:flux}
\end{eqnarray}
where $f_{\alpha}$ is the fraction of binding energy going into the
Ly$\alpha$ line. Other model assumptions included: 1) The gas density 
profile was assumed to trace that of the dark matter at large radii, 
but with a finite core. The emissivity profile was calculated 
self--consistently by computing the rate of change in the gravitational 
binding energy at each radius; 2) The infall velocity of the gas was taken 
to be of the form $v_{\rm bulk}(r)=-v_{\rm amp}(r/r_{\rm vir})^{\alpha}$. 
The $\alpha$ dependence of $f_{{\rm Ly}\alpha}$ is a fit that is valid for 
the range $\alpha \in [-0.5,1]$ (see paper I for more details
on our models).

We caution that the strong break in the slope of the redshift dependence of the flux (eq.~\ref{eq:flux}) at $z = 4$ should not be taken literally. In reality, the variation of the this slope with redshift should be smoother. The main point of eq.~(\ref{eq:flux}) however, is that because the IGM becomes increasingly opaque towards higher redshifts at $z \gtrsim 4$, cooling radiation of a fixed luminosity is more difficult to detect at $z \gtrsim 4$ than at $z \lsim 4$. Quantitatively, the total detectable Ly$\alpha$ cooling flux does {\it not} scale with $d_L^{-2}$ (which can be approximated as $\propto (1+z)^{-2.75}$), but as $ \propto <e^{-\tau}>d_L^{-2}$ (which can be approximated as $\propto(1+z)^{-11}$) for $z \geq 4.0$.\footnote{This conclusion does not account for the proximity effect near bright QSOs, which may allow significantly more flux to be transmitted (see Paper I for more details).}

In paper I, we studied a range of models with different masses,
redshifts, and emissivity, density, and velocity
profiles. Table~\ref{table:models} summarizes the parameters of two of paper I's fiducial models and two models used in \S~\ref{sec:matsuda}. For each model
the table contains the total mass, redshift, and circular velocity of the
host halo, as well as the parameters describing the infall velocity
field. For each model, all gas continously emits all its gravitational binding energy as Ly$\alpha$ as it navigates down the dark matter potential well (in the terminology of paper I, these are 'extended' models with $f_{\alpha}=1.0$).
\begin{table}[ht]
\small
\caption{Model Parameters}
\label{table:models}
\begin{center}
\begin{tabular}{ccccccc}
\tablewidth{3in}
Model \# & $M_{\rm tot}$  & $z_{\rm vir}$ & $v_{\rm circ}$ &$v_{\rm amp} $ & $\alpha$ & $f_{\alpha}$\\
(Ext.) & $10^{11}M_{\odot}$  & & km s$^{-1}$ & $v_{\rm circ}$  & & \\
\hline
\hline
1 & 5.2 &3.0 & 181&1.0 &1.0 &1.0\\
2 & 5.2 &3.0 & 181&1.0 &-0.5 &1.0\\
M1 & 15 & 3.0 & 257&1.0 &1.0&1.0\\
M2 & 15 & 3.0 & 257&1.0 &-0.5&1.0\\
\hline
\hline
\end{tabular}\\[12pt]
\end{center}
\end{table}

\subsection{Comparison with Steidel et al's LABs.}
\label{sec:steidel}

We start our comparison with the data presented in \citet{Steidel00},
who found two large, diffuse, and relatively bright Ly$\alpha$
emitters (`Ly$\alpha$ blobs' or 'LABs') in a deep narrow band search
of a field known to contain a significant overdensity of Lyman Break
galaxies. The total observed flux is $\sim 10^{-15}$ ergs s$^{-1}$
cm$^{-2}$ for each blob, and both have a physical size exceeding $140$
kpc. The excitation or ionization mechanism for either blob is poorly
understood. One blob, referred to as `LAB1', has a bright sub--mm
source as a counterpart \citep[with some uncertainty in the
identification, given the 15'' half power beamwidth;
][]{Chapman03}. The physical interpretation is that a dust enshrouded
AGN, which is undetected at other frequencies, is buried within the
Ly$\alpha$ halo. It has been speculated that a jet--like structure
that is well outside of our line of sight induces star formation,
which, combined with the ionizing radiation from the AGN,
photo-ionizes the surrounding hydrogen cloud \citep{Chapman04}. The
other blob, `LAB2', has no sub-mm bright counterpart, but
\citet{Antara04} have found an associated hard X-ray source in {\it
Chandra} archival data, with a total luminosity comparable to that of
the Ly$\alpha$ ($L \sim 10^{44}$ ergs s$^{-1}$). \citet{Antara04} have
speculated that both LAB1 and LAB2 may have an embedded AGN, which
goes through periodic episodes of activity. An old population of
relativistic electrons, originating near the supermassive black hole
that powers the AGN, can upscatter CMB photons into the X-ray band,
which may then illuminate the hydrogen gas and explain at least part
of the Ly$\alpha$ flux in these blobs. This scenario is based on
interpretations of recent findings of extended X-ray sources at
redshifts as high as $z=3.8$ \citep[e.g.][]{Scharf03}.

The possibility that cooling radiation contributes significantly to
Ly$\alpha$ flux of the LABs remains. First, we notice that to produce
a flux of $10^{-15}$ ergs s$^{-1}$ cm$^{-2}$ from cooling radiation
alone, requires $v_{\rm circ} \sim 300$ km s$^{-1}$ or $M_{\rm tot}
\sim 2-3 \times 10^{12} f_{\alpha}^{-3/5}\hs M_{\odot}$.  The virial
radius of an object of this mass is $\sim 100 f_{\alpha}^{-1/5}$ kpc,
corresponding to $\sim 13$ $f_{\alpha}^{-1/5}$ '', in good agreement
with the observed size of the blobs.  Given that the image can be
observed over a large region on the sky suggests that the surface
brightness profile is not declining rapidly with radius, which argue
against our ``central'' model. More quantitatively, Figure 9. of
paper I shows that $\partial \ln F$ / $\partial \ln \theta$ $\lsim -2$
and $\gsim -1$ for the central and extended models, respectively for
$M=4 \times 10^{12} {\rm M}_{\odot}$. The plots in \citet{Steidel00}
lack contour levels, but especially the surface brightness of fainter
emission does not appear to rapidly decrease with radius.

For extended cooling radiation, the Ly$\alpha$ spectrum should be
double peaked (provided the red peak is not swallowed by the blue, but
for $v_{\rm amp}\sim v_{\rm circ} \sim 300$ km s$^{-1}$, this is not
expected according to our findings in paper I). \citet{Steidel00}
present spectra taken along slits across both blobs. More recently,
however, \citet{Bower04} and \citet{Wilman05} have used the SAURON
Integral Field Spectrograph to observe \citet{Steidel00}'s LAB1 and
LAB2, respectively. Since these spectra cover the entire
two--dimensional region of Ly$\alpha$ emission, we use these more
recent spectra for comparisons to our models.

\subsubsection{Comparison with Spectra of LAB1}
\label{sec:LAB1}
The spectra presented by
\citet{Bower04} have a complex structure. If LAB1 is emitting cooling
radiation, surrounded by an extremely optically thick ($\tau\gsim
10^6$) collapsing hydrogen cloud, the Ly$\alpha$ radiation is expected
to emerge with a net blue shift. It is not possible to tell from
\citet{Bower04}'s data whether this is the case, since, in the absence
of other lines, the systemic redshift of the LAB is not known. There
is no evidence for a fainter redder peak. In \citet{Steidel00}'s
original data, there {\it is} evidence that the main bluer peak is
accompanied by a fainter redder peak, with $[B]/[R]\gg 1$.
Unfortunately, it is not possible to calculate the exact value of
$[B]/[R]$ from \citet{Steidel00}'s paper, since the contour levels of
their spectra are not shown. In our fiducial model, but with the mass
of the halo increased to $M_{\rm tot} = 3 \times 10^{12} {\rm
M_{\odot}}$, as required by the angular size and total flux from LAB1,
we expect $[B]/[R]\sim 100$ after including the effect of the IGM
opacity (see Figure 8 in paper I for how the [B]/[R] ratio scales
with various parameters).  A secure detection of a double--peaked,
asymmetric profile, with a stronger blue peak, $[B]/[R]\gg 1$, would
be strong evidence for an optically thick, collapsing gas cloud.
\citet{Bower04} note that the line width appears to increase with
radius, a feature that does not arise in models of neutral collapsing
gas clouds presented in paper I. However, it does arise in models in
which an ionized source is embedded within the gas cloud (see Fig. 13
of paper I). Indeed, the presence of a sub--mm source suggests that an
ionizing source resides within LAB1. The ionizing source can not,
however, fully ionize the gas (Fig. 13 in paper I), without violating
$[B]/[R]\gg 1$ as observed by \citet{Steidel00}.

The origin of the \lya emission in our model is then attributed to
fluorescent radiation emitted in an HII region created by the central
ionizing source. This HII region is surrounded by a neutral collapsing
hydrogen envelope. Cooling radiation, generated over a spatially
extended region within the neutral hydrogen envelope, may provide an
additional contribution to the \lya luminosity. Evidence for this
scenario would be (i) a flat surface brightness profile, (ii) the
strong asymmetry of the double--peaked profile ($[B]/[R]\gg 1$) and
(iii) the increase of the FWHM of the line with radius. This interpretation
is consistent with that given by \citet{Bower04}.

We note that \citet{Ohyama03} have also obtained deep spectroscopy in
a slit across LAB1, and showed that the spectrum of the brightest knot
in particular, is more complex than can be explained with our
models. The spectrum of the central knot can be be fitted with three
Gaussians, each separated by $\sim 2000$ km s$^{-1}$. The other
regions are consistent with our model above.\\

\subsubsection{Comparison with Spectra of LAB2}
\label{sec:LAB2}
LAB2 has a complicated
morphology, with six distinct sub--blobs of resolved Ly$\alpha$
emission. The spectra of the sub--blobs all exhibit a double--peaked
frequency structure. \citet{Wilman05} show that the spectra can be
explained with a very simple model. The emission line is assumed to be
a Gaussian with a FWHM of $\sim 1000$ km s$^{-1}$, with its peak
frequency varying by $\sim 290$ km s$^{-1}$ over the entire LAB. A
foreground slab of neutral gas with $N_{HI} \sim 10^{19}$ cm$^{-2}$
leaves an absorption feature which falls at the same wavelength in all
six spectra. \citet{Wilman05} attribute this absorption feature to a
foreground shell of neutral hydrogen gas, swept up by a galaxy--sized
outflow. Indeed, in four out of six sub--blobs (except for B and F),
the absorption line lies blueward of the peak of their Gaussian
emission line.  We show below, however, that these data can also be
explained with an equally simple in--fall model.

\begin{figure}[ht]
\vbox{ \centerline{\epsfig{file=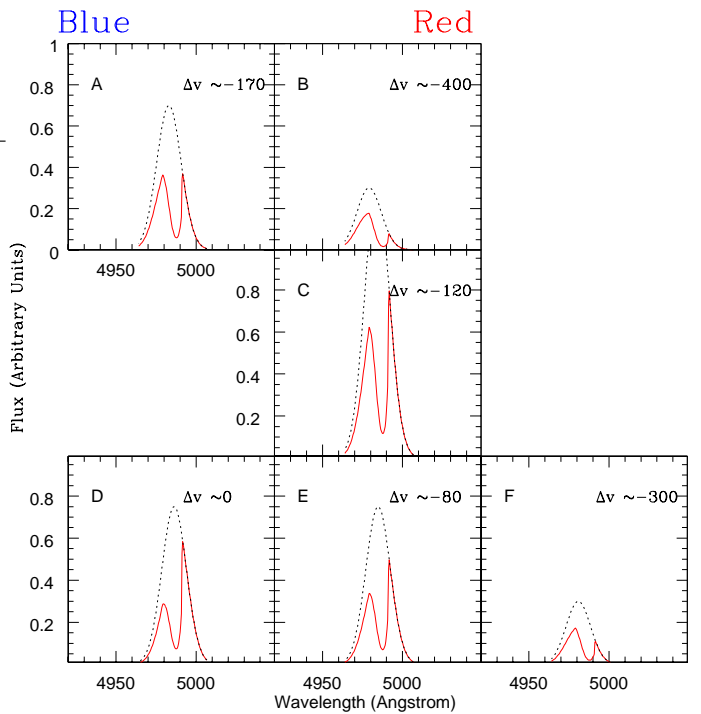,width=10.0truecm}}}
\caption[The Detection of IGM Infall around Steidel et al 2000's
LAB2]{A reproduction of the observed Ly$\alpha$ spectra of 
Steidel et al's LAB2 by \citet{Wilman05}. The spectra 
in the different panels are displayed in the same way 
as originally in \citet{Wilman05}, who succesfully reproduced
their observations with a foreground shell of neutral hydrogen gas, 
swept up by a galaxy--sized outflow. This figure demonstrates
 that our infall model reproduces the observed spectra of LAB2 equally
 well. Our model assumes simple infall for the
IGM, as outlined in \S~\ref{sec:IGM}. The {\it red-solid lines} are
the spectra after scattering by the IGM. The {\it solid-black lines}
are the intrinsic spectra. The spectra in panels B and F,
with enhanced blue peaks, are consistent with cooling radiation that
is propagating outwards through the collapsing, neutral hydrogen gas.}
\label{fig:wilman} 
\end{figure}

In our model, we take the same intrinsic emission profile of the
Ly$\alpha$ emission as in \citet{Wilman05}, and likewise, we allow
ourselves the freedom to shift the peak frequency of the intrinsic
emission to the blue by up to $\sim 300$ km s$^{-1}$ (see the shift
assumed in each panel). We process this emission line through the
model of the dynamic IGM as discussed in \S~\ref{sec:IGM}.  We
calculated the IGM transmission curves expected for a source with
$M_{\rm tot}=10^{12}M_{\odot}$ and added an ionizing source with
$L_{\rm ion}=2 \times 10^{44}$ ergs s$^{-1}$ to increase the
transmission near the line center (Note that we assumed $\theta=0$ in
eq.~\ref{eq:tauigm} for all panels, as discussed in
\S~\ref{sec:IGM}). The results are shown in Figure~\ref{fig:wilman},
which can be compared directly to Figure 3 in \citet{Wilman05}. Our
model provides an equally good fit to the data. Further improvement of
the fits is possible, if we allow ourselves the freedom to vary the
IGM model (and the exact shape of the emission line) from location to
location on the sky, rather than imposing a single IGM infall model as
we have done.

To understand why our model provides a good fit is through the IGM transmission curve shown in {\it lower left panel} of Figure~\ref{fig:igmtransmission}\footnote{We point out that in Figure~\ref{fig:wilman} the horizontal axis displays the observable Ly$\alpha$ wavelength (blue and red lies on the left and
right, respectively), while the horizontal axis of
Figure~\ref{fig:igmtransmission} displays a frequency variable (and,
consequently, blue and red lies on the right and left,
respectively).}. Panel D in Figure~\ref{fig:wilman} is obtained by 
multiplying a Gaussian emission line that is centered on the line center ($x=0$), by this IGM transmission. Shifting the Gaussian back and
forth in frequency and multiplying these emission lines by this
transmission curve yields the results shown in
Figure~\ref{fig:wilman}. Each panel contains the required velocity
shift in its top right corner. A negative velocity corresponds to a
blue shift of the line. \\

First we point out that all panels, except for panel D, require an
intrinsic blue shift of the emission line. This requires the gas
within the virial radius to be collapsing. Second, the spectra in
panels B and F, which are the two farthest out--lying sub--blobs in
the two--dimensional image of LAB2, have enhanced blue peaks,
requiring a significant blueshift in the intrinsic emission, by $\sim
3-400$ km s$^{-1}$.  Within our model, this blue--shift would
naturally arise if the gas far out at the locations of sub--blobs B
and F is collapsing and optically thick ($\tau_\alpha \gtrsim 10^4$)
to Ly$\alpha$ photons. The smaller intrinsic required blueshift in
panels A, C and E can be accounted for if the Ly$\alpha$ photons were
propagating outwards through gas that is photoionized by an embedded
ionizing source (see Figure 11 in paper I). \citet{Wilman05} show in
their Figure~2 that regions B and F are also significantly fainter
than the other sub--blobs.  This would again be expected the \lya
radiation emerging from these regions were attributed to cooling
radiation (as opposed to photoionization for the other sub--blobs,
which reside closer to any embedded ionizing source). Our physical
picture associated with the collapse is that the gas is collapsing as
a whole onto their mutual center of mass. Panels B \& F show Ly$\alpha$
cooling radiation, generated over a spatially extended region.
The Ly$\alpha$ emission from the blobs in the other panels consists of processed ionizing radiation from embedded ionizing sources.
 The IGM surrounding LAB2 is dragged along in the overall cloud collapse.\\

Based on the data on the two Ly$\alpha$ blobs presented in
\citet{Steidel00}, we conclude that the spectrum and surface
brightness distribution of LAB1 are consistent with a model in which
gas is collapsing in a DM halo with $M_{\rm tot}=$ a few $\times
10^{12}M_{\odot}$ while emitting fluorescent Ly$\alpha$ radiation in
response to photoionization by (an) embedded ionizing source(s)
\citep[except the brightest knot, which is likely powered by an
outflow][]{Ohyama03}.  We found that the spectra of LAB2 can be
explained if the intrinsic Ly$\alpha$ emission profile is a single
Gaussian, blue--shifted by up to several hundred km/s due the collapse
of optically thick gas within a massive halo, and subsequent resonant
scattering on larger scales within a perturbed IGM.

\subsection{Comparison with Matsuda et al.'s LABs}
\label{sec:matsuda}

More recently, \citet{matsuda04} published a catalogue of an
additional 33 LABs, discovered in a deep, narrow--band, wide field
(31' $\times$ 23') search with {\it Subaru}, centered on the same
field as studied by \citet{Steidel00}.  One third of their blobs are
not associated with bright UV sources, and thus provide good
candidates for objects dominated by cooling radiation. \citet{Mori04}
have used ultra--high resolution simulations to model a forming galaxy
undergoing multiple epochs of supernova explosions and found their
models could reproduce the typical Ly$\alpha$ luminosities observed
from these LABs. Their models are particularly successful in
reproducing the 'bubbly' appearance of several LABs. \citet{matsuda04}
suggest that five specific Ly$\alpha$ blobs may be cooling objects,
based on the diffuse appearance of the images. A diffuse appearance,
may not be a necessary requirement for cooling radiation. The physical
picture associated with extended cooling is that clumps of cold gas
are flowing inwards, while emitting Ly$\alpha$ in response to various
sources of heating (i.e. gravitational shock heating and possibly
photoionization heating by the thermal emission from the hot, ambient
gas). Therefore, some clumpiness in the surface brightness
distribution may be expected. If we drop the requirement that the
Ly$\alpha$ image is diffuse, then the number of candidates for cooling
objects is increased to $12$.

\begin{figure}[t]
\vbox{ \centerline{\epsfig{file=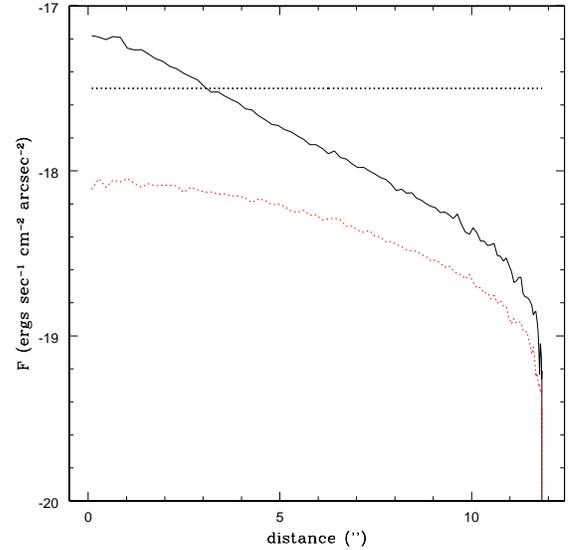,width=8.0truecm}}}
\caption[Surface brightness profiles of 2 models that fit
observations.]{The figure shows the surface brightness
profile of a model that has the observed gross properties of the
majority of the blobs found by \citet{matsuda04} ({\it black solid line}, 
see text). If this model is correct, then deeper observations will reveal these blobs to be spatially more extended. The {\it red dotted line} represents the
same model with $\alpha$ changed from $-0.5$ to $1.0$. This object
would escape detection since its surface brightness is everywhere
below \citet{matsuda04}'s detection limit, indicated by the {\it
thick--dotted horizontal line}.}
\label{fig:matsuda} 
\end{figure}

The detected fluxes for the majority of the LABs lie in the range $7
\times 10^{-17}-2 \times 10^{-16}$ ergs s$^{-1}$ cm$^{-2}$, which
correspond to cooling radiation from objects in the mass range $M_{\rm
tot}\sim 1-2$ $\times 10^{12} f_{\alpha}^{-3/5}M_{\odot}$, for
extended models with $(v_{\rm amp},\alpha)$ $=(v_{\rm circ},1)$ (see
eq.~\ref{eq:flux}). The corresponding range in virial radii is $\sim
80-100 f_{\alpha}^{-1/5}$ kpc, which translates to an angle on the sky
of $\sim 10-12 f_{\alpha}^{-1/5}$''. Converting this to an area yields
values which exceed the observed Ly$\alpha$ isophotal area by a factor
of $\sim 10$. This is not surprising since the calculated surface
brightness distribution may drop below the detection threshold in the
outer parts. This is illustrated in Figure~\ref{fig:matsuda}, in which
we plot the surface brightness profile for model M1 as the {\it
red--dotted line}. This is equivalent to our fiducial model, but with
$M_{\rm tot}$ increased to $1.5 \times 10^{12} M_{\odot}$.
 The total flux on earth for this model is
predicted to be $1.6 \times 10^{-16}$ ergs s$^{-1}$ cm$^{-2}$.

From Figure~\ref{fig:matsuda}, however, it can be seen that this
radiation is too diffuse to be detected. The total flux is spread over
an area of $\sim 300$ arcsec$^2$, which brings the average surface
brightness below the detection threshold of $3 \times 10^{-18}$ ergs
s$^{-1}$ cm $^{-2}$ arcsec$^{-2}$. The flux is more centrally
concentrated for models with lower $\alpha$ (see paper I). Plotted as
the {\it black--solid line}, is the surface brightness profile of
model M2, which is identical to model M1, except with
$\alpha=-0.5$. For this model, the flux from the inner 3'' exceeds the
detection limit. The total flux from this detectable region is $1.5
\times 10^{-16}$ ergs s$^{-1}$ cm$^{-2}$, which is spread over an area
of $\sim 30$ arcsec$^2$. This angular extent is consistent with the
observed area of the majority of Ly$\alpha$ blobs, $20-40$
arcsec$^2$. The average surface brightness of the detectable region of
model M2 is $\sim 5.2 \times 10^{-18}$ ergs s$^{-1}$ cm $^{-2}$
arcsec$^{-2}$, which corresponds to $m_{AB}\sim 27.1$ mag
arcsec$^{-2}$. \citet{matsuda04} focussed on imaging and did not
present spectra in their paper, so further comparison is at this stage
not possible. If this model is correct, then deeper observations will
reveal these sources to be spatially more extended. We showed in
Figure~10 of paper I that for extended models with $\alpha=-0.5$,
the bluest 15$\%$ of the Ly$\alpha$ photons would appear more
centrally concentrated than the reddest $15\%$. Comparing Figure~10
from paper I and Figure~\ref{fig:matsuda} suggests that the surface
brightness of the reddest $15\%$ of the photons may have been below
the detection limit, implying that deeper observations may reveal the
red peak in the spectrum (if it is not already in the data). 

\subsection{Spectroscopic Detections}
\label{sec:spectra}

Our comparisons have so far focused on extended LABs, which are a
rarity among the known Ly$\alpha$ emitters. As mentioned in
\S~\ref{sec:intro}, narrow band searches on {\it Kitt Peak}
\citep[LALA, e.g.][]{rhoads00,Rhoads01,Rhoads04}, on {\it Keck}
\citep[e.g.][]{Hu98} and on {\it Subaru}
\citep[e.g.][]{Kodaira03,Hu04,Ouchi05}, have been very successful in
finding galaxies at redshifts $z \geq 4.5$.  The narrow band surveys,
combined with deep broad band images, provide candidates based on
selection criteria involving the colors, equivalent widths, and flux
thresholds. These candidates need spectroscopic confirmation, using
large telescopes such as the {\it Very Large Telescope} (VLT) and {\it
Keck}. Roughly $30 \%$ of the candidates are confirmed to be genuine
high redshift Ly$\alpha$ emitters \citep[e.g.][]{Hu04}. Contrary to
the lower redshift LABs, these Ly$\alpha$ emitters thus all have
observed spectra. The known Ly$\alpha$ emitters typically have total
fluxes in the range $5\times 10^{-18}-5 \times 10^{-17}$ ergs s$^{-1}$
cm$^{-2}$, which is fainter by at least an order of magnitude than the
LABs discussed above.

If some of these Ly$\alpha$ emitters are cooling objects, then deeper
observations will reveal fainter, more diffuse emission associated
with them (unless the emission is strongly centrally concentrated, in
which case the surface brightness profiles drops rapidly, see Fig.~11
of paper I). To produce a detectable flux of $10^{-17}$ ergs s$^{-1}$
cm$^{-2}$ from a $z=5.7$ source requires $M_{\rm tot}\sim
10^{12}M_{\odot}$ (eq.~\ref{eq:flux}). A dark matter halo of this mass
that collapses at $z=5.7$ is associated with a $\sim 3\sigma$
fluctuation in the primordial density field. According to the
Press-Schechter mass function, the number density of halos with $M >
10^{12}M_{\odot}$, is $1.3 \times 10^{-5}$ Mpc$^{-3}$ at $z=5.7$,
which is factor of $\sim 40$ lower than the implied number density of
\lya emitters at this redshift \citep[e.g.][]{Ouchi05}. Additionally,
the typical observed flux from a Ly$\alpha$ emitter of $10^{-17}$ ergs
s$^{-1}$ cm$^{-2}$ comes from a small region, $\sim 1$
arcsec$^2$. This implies that either all cooling radiation from this
$M_{\rm tot}=10^{12}M_{\odot}$ object emerges from this small area,
which requires the surface brightness profile to be very steep, and
thus favor the central models (\S~\ref{sec:intro}). Alternatively, we
could be detecting the brightest central region a spatially extended
Ly$\alpha$ halo (as in Fig.~\ref{fig:matsuda}). This, however, would
require an even more massive, and therefore even rarer, halo to
produce the measured Ly$\alpha$ flux levels. The above suggests that
the majority of the known high redshift Ly$\alpha$ emitters are not
powered by cooling radiation alone. Indeed, the continuum detection of
many high redshift Ly$\alpha$ emitters, which show that the equivalent
width of the Ly$\alpha$ line is consistent with ionization by young
stars forming with a usual IMF.

An object of particular interest to us is described by
\citet{Bunker032}, who discovered a $z=5.78$ galaxy with a total
Ly$\alpha$ flux of $2 \times 10^{-17}$ ergs s$^{-1}$ cm$^{-2}$ in the
{\it Chandra Deep Field-South}. Using the DEIMOS spectrograph on {\it
Keck}, they obtained a spectrum, which looks remarkably like the
spectrum for a 'extended' model: a small red peak is separated from
a more pronounced bluer peak by $\sim 400$ km s$^{-1}$. If powered by
cooling radiation, a total flux of $2 \times 10^{-17}$ ergs s$^{-1}$
cm$^{-2}$ would require gas cooling in a halo with a mass of $M\sim
10^{12} M_{\odot}$. If the gas in this object were continuously
cooling, this would produce the maximum of the red peak to lie $>300$
km s$^{-1}$ from the line center. The problem with this interpretation
however, is that the Ly$\alpha$ emission is extremely compact, with an
angular size less than $\sim 0.1''$. Since the emission is detected at
the $20-\sigma$ level, this implies that the surface brightness falls
of by a factor of $\gsim 20$ over $\sim 0.1''$. Because extended
cooling models have rather flat surface brightness distributions (with
$\partial \ln F$ / $\partial \ln \theta$ $\sim -1$ to $-0.5$, see
paper I), they are not consistent with the observed compactness of the
emission. A similar conclusion applies to the faint, strongly
gravitationally lensed Ly$\alpha$ emitter discovered recently behind a
galaxy cluster \citep{Ellis01}, which remains a point--source despite
the large increase in the effective spatial resolution afforded by the
presence of strong lensing.

\begin{figure}[t]
\vbox{ \centerline{\epsfig{file=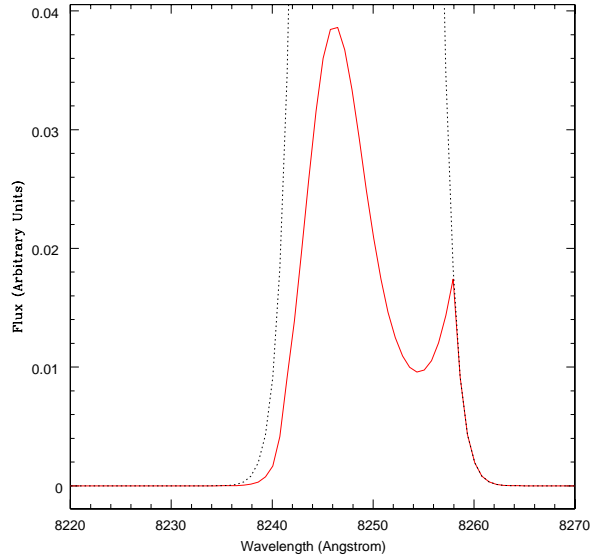,width=8.0truecm}}}
\caption[An Model for a known $z=5.78$ Emitter]{ This figure shows a
the spectrum of a model that is consistent with recent observations by
of a $z=5.78$ Ly$\alpha$ emitter in the {\it Chandra Deep
Field--South} by \citet{Bunker032}. An infalling IGM, as outlined in
\S~\ref{sec:IGM}, can modify the intrinsic Gaussian emission line
(indicated by the {\it black--dotted line}) to have a dip in the
observable spectrum (indicated by the {\it red--solid} line) on the
red side of the line.}
\label{fig:bunker} 
\end{figure}

An alternative interpretation of this spectrum combined with the
compactness of the emission is that the Ly$\alpha$ emission is due to
stars in an already formed galaxy, as \citet{Bunker032} argue, that
fully photoionizes the gas in the galaxy. The emission line is
therefore centered on the true line center and its FWHM set by the
gas' velocity dispersion. If we process this Ly$\alpha$ through an
infalling IGM, as described in \S~\ref{sec:IGM}, we find we can
reproduce the gross features of the observed spectrum, as shown in
Figure~\ref{fig:bunker} by the {\it solid--red line}. The
intrinsic line profile is given by the {\it black--dotted
line}. In this model, the IGM scatters $\sim 95 \%$ of the
Ly$\alpha$ photons out of the line of sight.  It is generally assumed
that the IGM only allows photons from high--z Ly$\alpha$ emitters to
be detected on the red half of the Ly$\alpha$ emission line, implying
$\sim 50\%$ is scattered out of our line of sight. If our model is
correct then the intrinsic luminosity may of this Ly$\alpha$ emitter
may have been underestimated by a factor of $\sim 10$.  The
intrinsic Ly$\alpha$ equivalent width would then be
increased from 20 to $\sim 200$ \AA. This could have important
consequences: if the intrinsic Ly$\alpha$ luminosity has been
systematically underestimated for high--z Ly$\alpha$ emitters, then
the escape fraction of Ly$\alpha$ photons would have to be unusually
high (for a constant star formation rate). This may put interesting
limits on the dust abundance and/or distribution in these
galaxies. Alternatively, it could imply that estimates of the total
intrinsic Ly$\alpha$ luminosity must be revised, requiring a $\sim 10$
times larger star--formation rate. The latter would enhance the
implied contribution of high redshift Ly$\alpha$ emitters to the
ionizing background. Either way, these issues deserve more attention
and will be addressed in more detail elsewhere \citep{Let}. We
emphasize that the only requirements for these issues to be relevant
for a Ly$\alpha$ emitter are that i) its surrounding IGM is
collapsing, and ii) its intrinsic emission line is not redshifted
significantly ($v_{\rm shift}\gsim v_{\rm circ}$, but note this 
can easily be achieved in outflows).

\section{Discussion}
\label{sec:discussion}

As argued above, observational evidence exists for gas infall around both pointlike (\S~\ref{sec:spectra}) and extended (\S~\ref{sec:LAB2}) Ly$\alpha$ emitters. Below (\S~\ref{sec:outflowsteidel}) we compare our model for \citet{Steidel00}'s LAB2 with a recently proposed outflow model by \citet{Wilman05}. We discuss why two such different mechanisms can reproduce the observed spectra well. We argue in \S~\ref{sec:outflow} that the notion that infall models may produce similar Ly$\alpha$ spectral features as outflow models may be more generic\footnote{Note the similarity of the Ly$\alpha$ spectrum produced by outflow and infall models is mainly due to the IGM and is therefore unrelated to our result (iv) in \S~\ref{sec:intro}} and discuss its cosmological implications. In \S~\ref{sec:ha} we discuss how additional imaging in H$\alpha$ may help to resolve this issue.

\subsection{Infall vs. Outflow in Steidel et al's LAB2}
\label{sec:outflowsteidel}

It may come as a surprise that two completely different mechanisms
(infall vs. outflow) may reproduce the observed spectra of \citet{Steidel00}'s
LAB2. To produce a spectrum with $[B]/[R] < 1$ using a single absorbing
shell of material, the shell must absorb on the blue side 
of the emission line, and must therefore be outflowing. 
As mentioned above, an infall model naturally produces $[B]/[R] < 1$ (shown in
Fig.~\ref{fig:igmtransmission}). Note that the for the exact same intrinsic
emission line and the same $[B]/[R]$ ratio, the dip in the spectra for
the outflow and infall model lies blueward and redward of the 
true line center, respectively. This difference is in practice
difficult to see, especially since the intrinsic emission line is
unlikely to be Gaussian (see Fig.~\ref{fig:fid.igm} and paper I, in
which we calculated the exact shape of the emission line emerging from
models of both neutral and photoionized collapsing gas clouds).
We caution that the isotropically infalling IGM will not
change the total flux toward Earth. However, the scattering 
"IGM shell" is geometrically quite thick (up to $\sim 5$ virial radii ),
 so any photons scattered back into our line of sight will be spread 
over a large solid angle (and also spread in frequency). This "super-fuzz" 
is too diffuse to modify the simple absorption dip due 
to scattering out of the line of sight.

An advantage of our explanation is that the intrinsic blue shift of
the emission line, in all panels, is consistent with overall gas
collapse. The amount of blueshift of each line is determined by the
brightness of the photoionized source. Indeed, the brightest
Ly$\alpha$ blobs have the smallest net blue shift.

A second advantage of the above explanation of the data is that it
may require less fine tuning than the explanation by the swept--up shell,
given by \citet{Wilman05}. The peak frequency of the absorption dip
varies by less than $60$ km s$^{-1}$ among the six different
sub--blobs covering LAB2. This small variation would arise naturally
if the entire shell has virtually stalled and come to rest with respect 
to the galaxy's systemic velocity. Because a superwind will spend most of its time close to the zero velocity stalling radius it is most likely observed 
at this position. However, for the shell to stall and preserve 
a similar column density over large length scales (a few $100$ kpc) 
and long times ($10^8$ yrs), requires the ambient pressure of the IGM
not to exhibit pressure gradients on these scales.
In our model, the IGM infall velocity is $\sim 300$ km s$^{-1}$ and may
vary by $\sim 20\%$ over the entire blob and may be a less stringent 
requirement. 

In either model, the difference between LAB1 and LAB2 remains a
puzzle. If the shape of the observed spectrum in LAB2 is indeed due
to IGM infall, then this IGM infall is also expected around
LAB1. However, no evidence for this is seen in the spectra of in LAB1.
In our infall model, this would require that the Ly$\alpha$ flux
emerges across LAB1 with a larger intrinsic blueshift of $\sim 500$ km
s$^{-1}$. As we showed in paper I, this can be easily reached when the
gas in LAB1 is optically thick ($\tau_\alpha \gsim 10^4$) to the
Ly$\alpha$ photons.

\subsection{Infall vs. Outflow in all Ly$\alpha$ Emitters}
\label{sec:outflow}

Ample spectroscopic evidence exist for outflows of gas around 
Lyman Break Galaxies \citep[LBGs, e.g.][and references therein]{Adelberger03}. 
It is not yet securely established to what level these 
superwinds surround the galaxy (i.e. what solid angle around the galaxy 
is swept up by the wind). The
physical extent of the superwind is set by the star formation history
of its host galaxy, and determines to what degree it can prevent gas
from the IGM to continue to cool and collapse onto the
galaxy. Understanding superwinds is clearly a requirement for
understanding galaxy formation and evolution. \citet{Wilman05} have
argued that their recent observations demonstrate that superwinds
occur on galactic scales around \citet{Steidel00}'s LAB2. However, we
showed in \S~\ref{sec:steidel} above that these observations are also
consistent with an inflowing IGM.

One may argue that the number of LBGs with known outflows greatly
outnumber the number of known inflows. However, this does not
necessarily imply that galaxies with outflows are more common than
those with significant gas infall.  At high redshifts, $z \gtrsim 5$,
candidates for galaxies containing a superwind are found based on
their Ly$\alpha$ properties \citep[see][for a review]{Taniguchi03}.
As in the case of LAB2, some of the Ly$\alpha$ spectra of candidate
superwind galaxies \citep[e.g.][]{Dawson02} could in fact also be
explained by infall of the IGM (Dijkstra et al, 2006b, in prep)
 onto the Ly$\alpha$
emitter. Additionally, for a given intrinsic Ly$\alpha$ luminosity,
there is observational bias to detect an outflow. The intrinsically
redshifted \lya emission from outflows is easier to detect than the
blueshifted Ly$\alpha$ from infall, because the latter will be subject
to much more severe attenuation in the IGM.  Furthermore, outflows can
be more compact than large scale infall, increasing the surface
brightness. These observational biases must be accounted for when
calculating the fraction of Ly$\alpha$ emitters associated with
outflows and infall. With ongoing and future Ly$\alpha$ surveys, the
sample of high redshift Ly$\alpha$ emitters steadily grows, allowing a
better determination of this number, and addressing various basic
questions related to galaxy formation: (i) how common is Ly$\alpha$
cooling radiation among collapsing protogalaxies? (ii) how common is
fluorescent Ly$\alpha$ emission due to embedded ionizing sources (such
as a central quasar or young stars that formed during the collapse)
among protogalaxies? iii) how common are superwinds in young galaxies
at high redshifts?

\subsection{H$\alpha$ Emission from Protogalaxies }
\label{sec:ha}

Discussion of emission line radiation from protogalaxies have focused
typically on Ly$\alpha$ radiation, but as argued by \citet{Oh99}, the
H$\alpha$ line may be a cleaner probe of the ionizing emissivity. The
production rate of \ha photons is $0.45$ per recombination, which is
comparable to that of Ly$\alpha$. Because hydrogen atoms spend such a
short time in their $n=2$ state, the \ha photons do not undergo
resonant scattering, and escape from protogalaxies in a single
flight. Because an \ha photon is $5/27$ times as energetic as
Ly$\alpha$, the total intrinsic \ha flux is lower. However, since \ha
photons do not suffer any attenuation in the IGM \citep{Oh99},
the detectable \ha flux can exceed that of Ly$\alpha$, especially for
neutral collapsing gas clouds at higher redshifts. A simultaneous detection of
both H$\alpha$ and Ly$\alpha$ would allow a separation of the intrinsic emissivity from radiative transfer effects. For example, a relative red or blueshift
of the Ly$\alpha$ line compared to the H$\alpha$ line is strongly indicative for an outflow or infall, respectively and separates infall from outflows.

Assuming that the intrinsic \ha flux is lower than the
Ly$\alpha$ flux by factor of $\mathcal{F}_{H\alpha} \sim 27/5=5.4$, we
estimate the total detectable H$\alpha$ flux emerging from a neutral
collapsing gas cloud (eq.~\ref{eq:flux}) to be:
\begin{equation}
\frac{f_{H\alpha}}{10^{-18}}\sim 1.7\Big{(}
\frac{2-\alpha}{1.75}\Big{)}^{1.2}\Big{(}
\frac{5}{\mathcal{F}_{H\alpha}} \Big{)} \Big{(}\frac{v_{\rm circ}}{116
\hs {\rm km}\hs {\rm s}^{-1} }\Big{)}^5
\Big{(}\frac{5}{1+z}\Big{)}^{2.75} \hs \frac {{\rm ergs}}{{\rm s}\hs
{\rm cm}^2},
\label{eq:haflux}
\end{equation} 
where we used $v_{\rm amp}=v_{\rm circ}$ and that
$f_{H\alpha}=L_{H\alpha}/[4\pi d^2_L]$.\footnote{We approximated
the luminosity distance by $d_L(z)$=$3.6\times 10^4$ $[(1+z)/5]^{11/8}$ Mpc
as in \S~\ref{sec:observations}.} 
Fluxes at this level will be detectable by the {\it James Webb
Space Telescope (JWST)}. The Near Infrared Spectrograph (NIRspec) on
{\it JWST} can detect a $\sim 100$ nJy point source at the $10 \sigma$
level in a $10^5$ s exposure time
\footnote{See http://www.stsci.edu/jwst/science/sensitivity.}. The
flux calculated in eq.~(\ref{eq:haflux}) is spread over a frequency
range of $\Delta \nu$ $\sim 2v_{\rm circ} \nu_{H\alpha}(1+z)/c$.  This
translates to a flux density of
\begin{equation}
\frac{f_{H\alpha,\nu}}{1400}\sim
 \Big{(}\frac{2-\alpha}{1.75}\Big{)}^{1.2} \Big{(}
 \frac{5}{\mathcal{F}_{H\alpha}}\Big{)} \Big{(}\frac{v_{\rm circ}}{116
 \hs {\rm km}\hs {\rm s}^{-1} }\Big{)}^2
 \Big{(}\frac{5}{1+z}\Big{)}^{1.75} \hs {\rm nJy},
\label{eq:hafluxden}
\end{equation} 
which is spread over an area $\pi \theta_{\rm vir}^2$, where
$\theta_{\rm vir}=r_{\rm vir}/d_A(z)$, in which $d_A$ is the angular
diameter distance. Given that a point source in NIRspec subtends only
$\sim 0.01$ arcsec$^{2}$ on the sky, the total noise over the area
over which the $H\alpha$ is emitted is $[(\pi \theta_{\rm
vir}^2)/(0.01$ arcsec$^2)]^{0.5}$ times higher. This translates to the
following signal to noise ratio in a $10^5$ s exposure:

\begin{eqnarray}
\frac{S}{N}\approx 10 \times \Big{(}\frac {f_{H\alpha,\nu}}{100\hs{\rm
    nJy}}\Big{)}\Big{(} \frac{0.01\hs{\rm arcsec}^2}{\pi \theta^2_{\rm
    vir}}\Big{)}^{1/2}\approx \nonumber \\ 2.6 \hs \sigma \Big{(}
\frac{2-\alpha}{1.75}\Big{)}^{1.2}
\Big{(}\frac{1+z}{5}\Big{)}^{-7/8}\Big{(}\frac{v_{\rm circ}}{116 {\rm
    km}\hs{\rm
    s}^{-1}}\Big{)}\Big{(}\frac{5}{\mathcal{F}_{H\alpha}}\Big{)},
\label{eq:tint}
\end{eqnarray} 
which is a very weak signal. However, the \ha emission is not
uniformly emerging from the area enclosed by $\theta_{\rm vir}$, but
is (much) higher in the central region of the image, which can boost
the expected flux densities. For model {\bf 1.} and {\bf 2.} (Table~\ref{table:models}) for example, $50\%$ of the flux comes from $\theta \lsim 0.27 \theta_{\rm vir}$ and $\theta \lsim 0.18 \theta_{\rm vir}$, respectively. This
would boost the $S/N$ for model {\bf 1.} and {\bf 2.} by a factor of
$\sim 2$ and $2.5$, respectively. It should be noted that with the
current proposed spectral resolution of $R=1000$, the \ha line would
not or barely be resolved. Because of increased detector noise at
$\lambda \gtrsim 5$ $\mu$m, the $S/N$ ratio drops by almost an order
of magnitude for H$\alpha$ at $z\gtrsim 7$. \citet{OhHaiman01} argued
that because the helium equivalent of H$\alpha$, which has a
transition at $\lambda=1640$ \AA, does not suffer from the increased
detector noise for $z \lsim 32$, it may be more easily
detectable. This would only apply to our self-ionized cases, and under
the restriction that the spectrum of the ionizing sources is
sufficiently hard $\beta \sim 1$ \citep[see][]{OhHaiman01}.

\section{Summary and Conclusions}
\label{sec:conclusions}

In paper I, we calculated the properties of the Ly$\alpha$ radiation
emerging from collapsing protogalaxies. In this paper we provided 
simple calculations of the mean transmission of the IGM around such objects
 to facilitate a comparison with observations. We assumed that the IGM surrounding our objects to be flowing in, following the recent prescription by Barkana (\S~\ref{sec:IGM}).  We found the main results from paper I
not to be affected significantly.

In \S~\ref{sec:observations} we compare our models with the
observations of known Ly$\alpha$ blobs presented by \citet{Steidel00}
in \S~\ref{sec:steidel}, and \citet{matsuda04} in
\S~\ref{sec:matsuda}.
The observed properties of those LABs without significant continuum in the larger sample of blobs presented by \citet{matsuda04}, such as the
surface brightness and angular size, are consistent with model in
which gas is cooling and condensing in dark matter halos with $M_{\rm
tot} \sim 1-2 \times 10^{12}M_{\odot}$. According to this model,
deeper observations should reveal that these source are spatially
extended by a factor of several beyond their currently imaged sizes
(see Fig.~\ref{fig:matsuda}). With the current data it may already be
possible to distinguish that the bluest 15$\%$ of the Ly$\alpha$
photons would appear more centrally concentrated than the reddest
$15\%$, which we propose as a diagnostic of optically thick gas
infall.

Apart from the brightest central knot, which
\citet{Ohyama03} have shown to be likely powered by an outflow, the
observed spectrum and surface brightness distribution of \citet{Steidel00}'s 
LAB1 are consistent with a model in which gas is collapsing in a DM halo with
$M_{\rm tot}=$ a few $\times 10^{12}M_{\odot}$. An embedded ionizing
source photoionizes the central part of the collapsing cloud, in which
fluorescent Ly$\alpha$ is emitted. Cooling radiation, emitted in the
surrounding neutral hydrogen envelope, may provide an additional
contribution to the Ly$\alpha$ luminosity. We found that the spectra
of LAB2, as presented recently by \citet{Wilman05}, can be explained
if the intrinsic Ly$\alpha$ emission profile is a single Gaussian,
blue--shifted by up to several hundred km/s due the collapse of
optically thick gas within a massive halo, combined with subsequent
resonant scattering on larger scales within a perturbed, infalling
IGM.
 
We found similar evidence for infall of the IGM onto the $z=5.78$
Ly$\alpha$ emitter observed by \citet{Bunker032} in its spectrum (\S~\ref{sec:spectra}). We argue that IGM infall around high redshift Ly$\alpha$ emitters 
could be a more common phenomenon than previously believed, as infall models may produce spectra similar to those of outflow models. If this is correct, then the intrinsic Ly$\alpha$ luminosities, and derived quantities, such as the star--formation rate of high redshift Ly$\alpha$ emitters, may have been underestimated significantly (\S~\ref{sec:outflow}).

Because of the increasing opacity of the IGM with redshift, we expect
the detectable Ly$\alpha$ fluxes from high redshift ($z \gtrsim 4.0$)
collapsing protogalaxies to be very weak, unless the intrinsic
Ly$\alpha$ luminosity is very high. We concluded that the majority of
the observed high redshift Ly$\alpha$ emitters (\S~\ref{sec:spectra})
are most likely not associated with collapsing protogalaxies. At high
redshift ($z \gtrsim 4.0$) the detectable H$\alpha$ flux emerging from
neutral collapsing gas clouds may dominate that of Ly$\alpha$
(\S~\ref{sec:ha}). In cases in which both the \lya and \ha are
detected simultaneously, it would be possible to separate the
intrinsic emissivity from radiative transfer effects.

The work in this paper emphasizes the importance of the Ly$\alpha$
line as a probe of the earliest stages of galaxy formation. In
particular, with the growing sample of high redshift Ly$\alpha$
emitters (both point like and extended), and with deeper observations,
our hope is that it will be possible to discover evidence for
significant gas infall in proto--galaxies, caught in the process of
their initial assembly.\\

MD thanks the Kapteyn Astronomical Institute, and ZH thanks the
E\"otv\"os University in Budapest, where part of this work was done,
for their hospitality.  ZH gratefully acknowledges partial support by
the National Science Foundation through grants AST-0307291 and
AST-0307200, by NASA through grants NNG 04GI88G and NNG 05GF14G, and by
the Hungarian Ministry of Education through a Gy\"orgy B\'ek\'esy
Fellowship. We thank the anonymous referee for his or her comments that improved the presentation of both our papers.


\end{document}